\documentclass[conference]{IEEEtran}

\usepackage{hyperref}
\usepackage{graphicx}
\usepackage[utf8]{inputenc}
\usepackage{array}
\usepackage{ragged2e}
\usepackage{tabularx}
\usepackage{tabularx}
\usepackage{subfig}
\usepackage{footmisc}
\usepackage{amsmath}
\usepackage{microtype}
\usepackage{textcase}
\usepackage{paralist}

\usepackage{caption}
\newcommand{\tableheadline}[1]{\textsc{#1}}

\newcommand{\ex}[1]{\emph{#1}}
\newcommand{\cat}[1]{\textsc{#1}}

\newcommand{\bottomrule}{\hline}
\newcommand{\toprule}{\hline}
\newcommand{\midrule}{\hline}
\newcolumntype{Z}{>{\raggedright\let\newline\\\arraybackslash}X}
\newcolumntype{P}[1]{>{\centering\arraybackslash}p{#1}}
\newcolumntype{R}[1]{>{\raggedleft\arraybackslash}p{#1}}
\ifCLASSINFOpdf
 
\else
  
\fi

\begin{document}
\title{Estimating the Dissemination of Social and Mobile Search in
Categories of Information Needs Using Websites as Proxies}

\author{\IEEEauthorblockN{Christoph Fuchs\IEEEauthorrefmark{1},
Akash Nayyar\IEEEauthorrefmark{2}, Ruth Nussbaumer\IEEEauthorrefmark{3}, and
Georg Groh\IEEEauthorrefmark{4}}
\IEEEauthorblockA{Department of Informatics\\TU M{\"u}nchen\\Garching bei M{\"u}nchen, Germany\\
Email: \IEEEauthorrefmark{1}fuchs@cs.tum.edu,
\IEEEauthorrefmark{2}akash.123srm@gmail.com,
\IEEEauthorrefmark{3}nussbaumer.ruth@icloud.com,
\IEEEauthorrefmark{4}grohg@cs.tum.edu}}
\maketitle

\begin{abstract}
With the increasing popularity of social means to satisfy information needs using Social Media (e.g., Social Media Question Asking, SMQA) or Social Information Retrieval approaches, this paper tries to identify types of information needs which are inherently social and therefore better suited for those techniques. We describe an experiment where prominent websites from various content categories are used to represent their respective content area and allow to correlate attributes of the content areas. The underlying assumption is that successful websites for focused content areas perfectly align with the information seekers' requirements when satisfying information needs in the respective content areas. Based on a manually collected dataset of URLs from websites covering a broad range of topics taken from Alexa\footnote{\url{http://www.alexa.com} (retrieved 2015-11-04)} (a company that publishes statistics about web traffic), a crowdsourcing approach is employed to rate the information needs that could get solved by the respective URLs according to several dimensions (incl. sociality and mobility) to investigate possible correlations with other attributes. Our results suggest that information needs which do not require a certain formal expertise play an important role in social information retrieval and that some content areas are better suited for social information retrieval (e.g., \cat{Factual Knowledge \& News}, \cat{Games}, \cat{Lifestyle}) than others (e.g., \cat{Health \& Lifestyle}).
 \end{abstract}

\IEEEpeerreviewmaketitle

\section{Introduction}

Asking others for help is a fundamental, most basic information retrieval behavior of humankind \cite{Horowitz2010,Mansilla2013}. With the rise of social media, additional input factors became available to improve search. Today, a recent analysis\footnote{\url{http://fortune.com/2015/08/18/facebook-google/} (retrieved 2016-01-15)} shows that the online networking platform Facebook leads more traffic to news sites than Google, the market-leading search engine provider. With an increasing number of available low-cost capabilities to sense the user's individual environment it becomes possible to grasp and consider the user's context in search situations. In the following experiment, we investigate whether information needs covering content areas with different characteristics benefit from context-awareness and/or social means to satisfy information needs. 

\section{Research Questions}

The objective of this experiment is to investigate whether there are types of information needs (either on a meta-level or content-wise) which are more ``mobile'' or ``social'' than others and whether specific attributes can predict or explain the ``sociality'' of an information need.

\section{Related Work}

\subsubsection{Search, Information Needs, and Relevance} 
\label{ssub:search_information_needs_and_relevance}

Mobile information needs have been investigated in many experiments so far. Church and Oliver \cite{Church2012} investigate mobile internet and search behavior, Kamvar et al. \cite{Kamvar2009,Kamvar2006} analyze usage patterns in mobile search scenarios for different types of devices (including mobile devices). Sohn et al. \cite{Sohn2008} examine mobile information needs in a diary study. Oeldorf-Hirsch et al. \cite{Hirsch2014} describe the results of a SMQA study, investigating which information needs got routed to social networks or search engines. Fuchs and Groh \cite{FuchsGroh2015} extend parts of this experiment and suggest that the information seeker's privacy might be the limiting factor for social means of information retrieval. McDonnell and Shiri \cite{McDonnell2011} present a comprehensive taxonomy for social search. Mizzaro \cite{Mizzaro1998} distinguishes between conscious and unconscious information needs and deduces respective definitions of relevance.


\subsubsection{Classification of Information Needs} 
\label{ssub:classification_of_information_needs}
Classifications for general information needs have been proposed by Spink et al. \cite{Sping2002}, Kamvar and Baluja \cite{Kamvar2006}, and Kamvar et al. \cite{Kamvar2009}. Church et al. \cite{Church2007} and Sohn et al. \cite{Sohn2008} present classifications for mobile information needs. For social information needs, Morris et al. \cite{Morris2010b} and Oeldorf-Hirsch et al. \cite{Hirsch2014} propose classifications. Dearman et al. \cite{Dearman2008} and Church et al. \cite{Church2012b} defined classification systems for mobile social information needs.

\section{Approach}

To structure information needs by content type, a content taxonomy provided by Alexa\footnote{\url{http://www.alexa.com} (retrieved 2015-11-04)}, a company that publishes information about web traffic (and is owned by Amazon), is used. Alexa also issues a list of the most popular websites for each category. Alexa estimates the popularity of a URL by tracking a subset of the web users with the Alexa toolbar, a plugin for web browsers. The approach to answer the research questions of this experiment is the following: the basic assumption is that the most prominent websites for each content category are successful because they satisfy the users' information needs (which brought the users to the website in the first place) in that specific content area in an adequate way. The successful websites offer functionalities and features that fit to the content area, so that people who engage with the content feel comfortable with it. Instead of analyzing the information needs that might occur in the different categories, we investigate the websites which satisfy the information needs in each content category (see next section), assuming that the websites ``respond'' to the queries in the most appropriate way.

\begin{figure}
\centering
\includegraphics[width=1\columnwidth, angle=0]{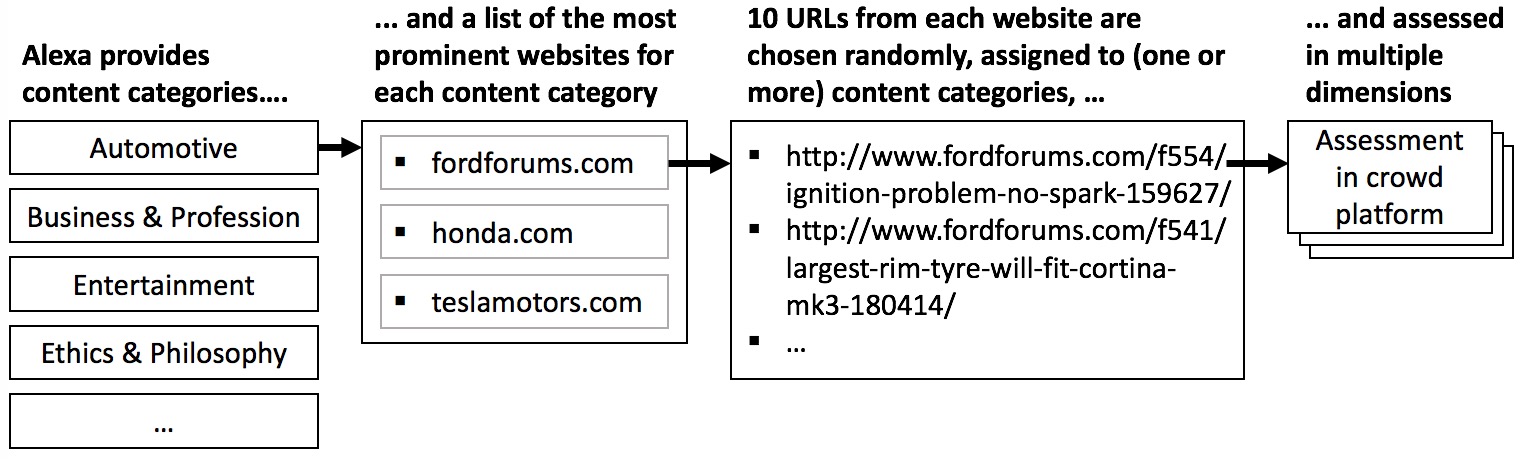}
\caption{Study approach}
\label{fig:exp3:approach}
\end{figure}

For each Alexa content category, the three most prominent websites have been selected (only exceptions: \cat{Travel} and \cat{Ethics \& Philosophy}, which only consist of two websites). Since some websites cover topics which can not be linked to a single content dimension, the topic of each website is expressed as a vector in Alexa's content category vector space holding percentage values for each content dimension. For each website, ten randomly chosen URLs have been selected and rated by participants of a web survey in different dimensions explained in the next section (\autoref{exp3:dataset}). \autoref{fig:exp3:approach} gives an overview of the approach. 
The participants of the survey received a small compensation for the task. We only accepted those judgments where the elapsed time between showing and submitting the survey form suggests that the user read and understood the questions (\autoref{exp3:data_collection}).

\section{Dataset}
\label{exp3:dataset}

Each URL was rated on several dimensions using a web survey. 
The dimensions can be divided in two groups: the first group (\autoref{ssub:axis_to_classify_information_needs}) is related to the hypothetical information need which would cause someone to visit the URL, the second group (\autoref{ssub:axis_to_classify_specialized_websites}) discusses the website's direct properties (business model, types of fostered social interaction, etc.). We do not claim that the dimensions are collectively exhaustive -- others do exist (cf. \autoref{exp3:limitations} for examples), but we focused on the ones listed in the next section because we assumed that those might have a high likelihood of showing differences between information needs with high and low levels of sociality.

\subsection{Dimensions to Classify Information Needs}
 \label{ssub:axis_to_classify_information_needs}
 
 In the following, the dimensions to classify the information need that is satisfied on the respective URL are explained in detail. 
 
 \subsubsection{Dependence on time} This dimension reflects whether the information need or the information have a certain expiration date or not. Possible values are
\begin{itemize}
	\item \cat{Hard Constraint} (2): The information need needs to be addressed at a specific point in time, e.g., if a user needs to get an idea for a good Christmas present, then the information need is clearly addressed at a specific time with the urgency depending on the current date.
	\item \cat{Soft Constraint} (1): The information need is addressed at a vague time, e.g., the information need is of the kind: \ex{``Any ideas for summer holidays?''}.
	\item \cat{Independent} (0): Not dependent on the time, e.g., the information need is of the kind: \ex{``What is your favorite football team?''}.
\end{itemize}

\subsubsection{Temporal validity} This dimension describes how long the information presented is valid. Possible values are
\begin{itemize}
	\item \cat{Long} (2): Reply to information need is valid for a very long time (e.g., decades, centuries, forever) , e.g., \ex{``When was Mozart born?''} -- the answer is valid forever.
	\item \cat{Medium} (1): Reply to information need is valid for a long time (e.g., months, maybe years), e.g., \ex{``Who is the current football player of the year?''}, or \ex{``How much does the new MacBook Pro cost?''}
	\item \cat{Short} (0): Reply to information need is valid for a short time (e.g. hours, days, maybe weeks). For instance, \ex{``Will it be sunny tomorrow?''}
\end{itemize}

\subsubsection{General applicability among users} This dimension describes to which degree the knowledge that is presented (and searched for in the hypothetical information need) is applicable for multiple users. Possible values are
\begin{itemize}
	\item \cat{High} (2): Information needs that are not tailored to one particular user; information need is shared by a lot of people, for instance: \ex{``What is the average cost of living in Munich?''}
	\item \cat{Medium} (1): Information needs that are important for a specific subset of people, for instance: \ex{``How to compute running time in Java?''}. In this case the given information need targets programmers.
	\item \cat{Low} (0): Information needs that are only important for a particular user, for instance: \ex{``My GRE score is 310. Is this sufficient to get admission in a good university in the United States?''}. In this case the information need is related only to the specific user who has a GRE score of 310.
\end{itemize}

\subsubsection{Knowledge codification} This dimension expresses to which degree the knowledge that is required to satisfy the information need is codified. In mature fields with commonly accepted explicit forms of knowledge representation (e.g., books, websites, etc.), knowledge is codified to a higher degree (e.g., medicine) than in areas where knowledge is widely discussed and controversial (e.g., user experience with the new BMW i3). Possible values are
\begin{itemize}
	\item \cat{High} (1): Knowledge to satisfy the information need is codified, i.e., it is defined in form of facts (books or articles) and there is a common agreement, e.g.: \ex{``What are the symptoms of Parkinson's disease?''}
	\item \cat{Low} (0): Knowledge to satisfy the information need is not codified, i.e., it is widely discussed and controversial; examples include questions asking for recommendation (e.g., \ex{``What is the best restaurant in Munich?''}) or opinions (e.g., \ex{``Do you like the new BMW i3?''})
\end{itemize}


\subsection{Dimensions to Classify Specialized Websites} 
\label{ssub:axis_to_classify_specialized_websites}

\subsubsection{Costs} The cost dimension describes whether the access to the information on the website is free or requires payment. Possible values are \cat{fee} (2), \cat{partially free} (1) and \cat{free} (0).

\subsubsection{Information provider} This attribute describes which profile fits best to the person who provides the information on the respective URL. Possible values are \cat{expert} (e.g., doctor, lawyer, editor), \cat{operator} (e.g., someone informing about own services or products), and \cat{layman} (i.e., someone who not necessarily has any formal expertise on the subject).

\subsubsection{Sociality} For each website, the existence of the following social features is evaluated and forms the degree of sociality (all features are weighted equally):
\begin{itemize}
	\item Can an ordinary user ask questions to satisfy an information need?
	\item Does the website recommend other content that was liked, commented on, viewed, or posted by others?
	\item Is there a possibility to rate or comment on the information need?
	\item Is there a possibility to create a personal profile?
	\item Is it possible to see what kind of information needs other people have or what kind of information needs they have satisfied before?
	\item Is it possible to contact the user who had the information need?
\end{itemize}

\subsubsection{Mobility} This dimension consists of three equally weighted sub-dimensions (aggregated to a single mobility value) and describes to which degree the underlying information need represents a ``typical'' mobile scenario. The participant is asked whether the information need which is satisfied by the given URL depends on a specific location. Valid answers include
\begin{itemize}
	\item \cat{High}: The user's physical location has a definite impact on the information need and the type of answer expected, e.g.: \ex{``Where is the ALDI supermarket closest to Klinikum Grosshadern metro station in Munich, Germany?''}
	\item \cat{Low}: The user's location does not impact the information need, for instance: \ex{``Which is the best Android phone in the market at present?''}
\end{itemize}

In addition, the participant is asked whether it is likely that the information need occurred in a mobile context and whether the information contains any specific spatial location information.



\subsection{Data Collection}
\label{exp3:data_collection}

Data collection was conducted using an online survey on a crowdsourcing platform\footnote{\url{https://microworkers.com/} (retrieved 2016-01-12)} with Indian participants. Each participant assessed ten randomly chosen URLs, using the dimensions outlined above. For each website, the data of the related URLs was aggregated using the average of the respective URL ratings and normalized on the interval $[0,1]$. To ensure data quality, all submissions that took less than $60$ seconds were excluded from the evaluation and were added to the pool of untreated URLs again. The threshold of $60$ seconds has been identified using test runs with skilled English speakers. In total, the dataset used for the analysis consists of $532$ evaluated URLs taken from $52$ websites.

\section{Results}
\subsection{Correlations}
The correlation of the different content categories and dimensions is shown in \autoref{tab:ex3_spearman} (Spearman's rho) and \autoref{tab:ex3_pearson} (Pearson's r). In the following, the findings will be briefly discussed for each dimension.

\subsubsection{Dependence on Time}
\cat{Dependence on Time} is positively correlated with \cat{Business Profession}, \cat{Automobiles}, \cat{Health \& Lifestyle}, and \cat{Entertainment}. While the first two content categories could possibly be explained by pressing information needs before (purchasing) decisions, the relation for the last two is less obvious.
The dimension \cat{Layman} is negatively correlated, which intuitively makes sense when considering that normal users will not be the best information providers when time critical information is requested. Content categories with a high negative correlation with dependence on time are \cat{Sports}, \cat{Games}, \cat{Real Estate}, and \cat{Society}. Especially for the categories \cat{Sports} and \cat{Real Estate}, this result is surprising. \cat{Real Estate} refers to renting or buying a property and the result might indicate that these decisions are rather short-dated than initially assumed. 

\subsubsection{Temporal Validity} 

Content in the category \cat{Ethics \& Philosophy} positively correlates with \cat{Temporal Validity}. This is not surprising, since the content is not expected to change fast. In contrast, content in the areas \cat{Entertainment}, \cat{Sports}, and \cat{Technology} varies at a much higher pace and therefore is negatively correlated.

\subsubsection{General Applicability}
On average, people's information needs regarding \cat{Travel} do not seem to differ much, since information in the \cat{Travel} category is positively correlated with \cat{General Applicability}. The findings suggest that the same applies to \cat{Business Profession} and \cat{Health \& Lifestyle}. In contrast, topics like \cat{Society}, \cat{Ethics \& Philosophy}, and \cat{Lifestyle} seem to be discussed quite individually -- for the \cat{Ethics \& Philosophy} category this comes a bit unforeseen, however, when taking the discussion and interpretation into account, the result may become more understandable.

\subsubsection{Knowledge Codification}
The findings suggest that \cat{Business Profession} and \cat{Recreation} tend to have a high degree of knowledge codification; in addition, it is also positively correlated with \cat{General Applicability} (Spearman's rho only) and \cat{Temporal Validity} (information that is valid for a long time or that is valid for a large group tend to be codified to a higher degree than other information). On the other end of the spectrum, information in the categories \cat{Society}, \cat{Games}, and \cat{Sports} are negatively correlated with \cat{Knowledge Codification}. The negative correlation with \cat{Factual Knowledge \& News} is surprising.

\subsubsection{Costs}
Websites in the content areas \cat{Lifestyles}, \cat{Technology} and \cat{Automobiles} tend to have higher results in the \cat{Costs} dimension. Also the categories \cat{Sociality} and \cat{Layman} are positively correlated with \cat{Costs}. The \cat{Costs} dimension is negatively correlated with \cat{Ethics \& Philosophy}, \cat{Health \& Lifestyles}, \cat{Trivia}, and \cat{Finance \& Insurance}. Especially the last category is unexpected because intuitively people would be willing to invest in serious topics like finance when stakes are high.

\subsubsection{Layman}
\cat{Layman} has a high positive correlation with \cat{Sociality}. In addition, it is positively correlated with topics in the categories \cat{Society} and \cat{Automobiles}. In contrast, \cat{Layman} is negatively correlated with \cat{Dependence on Time} and \cat{Operator}, which could be caused by the fact that laymen typically need some time to reply to information needs and the type of interaction. In addition, \cat{Layman} is negatively correlated with \cat{Entertainment}, which is surprising since one could intuitively assume that ``informal'' topics are related to less professional interaction modes. 

\subsubsection{Operator}
\cat{Operator} is positively correlated with the content categories \cat{Homes \& Garden}, and \cat{Sports}. A negative correlation exists for the dimensions \cat{Sociality}, \cat{Finance \& Insurance}, and \cat{Society}. 

\subsubsection{Expert}
The \cat{Expert} dimension is positively correlated with \cat{Automobiles}, \cat{General Applicability}, and \cat{Real Estate}. It seems valid to assume that the knowledge of experts is applicable to a larger audience and that expensive purchasing decisions might be backed up by acknowledged expertise. The positive correlation with \cat{Knowledge Codification} suggests that experts work in mature, clearly distinguished fields with commonly accepted methods and a documented state of the art. Negatively correlated are \cat{Sports}, \cat{Technology}, and \cat{Mobility}. The first two categories could be explained by the fact that both are content-driven and expertise might be easier to gain (or maybe difficult to get, because of low degree of knowledge codification as in \cat{Sports}). The negative correlation between \cat{Expert} and \cat{Mobility} could possibly be explained because experts for a certain spatial area are often not considered as ``professional'' experts and therefore correspond more with the \cat{Layman} category. 

\subsection{Explaining Sociality}
\label{exp3:res:explain_social}

Apart from general correlations of attributes (as discussed in the previous section), it is also interesting how \cat{Sociality} can be explained using the other factors as explanatory variables. A high degree of explanation could suggest that social interaction plays an important role in some specific content areas and that some attributes of information needs would encourage the use of social means to satisfy the information need (and vice versa, i.e. some information needs and content areas are not suited for social information retrieval). Therefore, a linear regression model was fitted based on the dimensions shown above. After applying an optimization using the Bayesian Information Criterion (BIC), the only factors kept are \cat{Dependence on Time}, \cat{Layman}, and \cat{Operator}. \autoref{tab:exp3_soc_model} shows the results for the linear model. The model's residuals are distributed normally to a sufficient degree (studentized Breusch-Pagan test: p-value = $0.36$, Goldfeld-Quandt test: p-value = $0.80$), and the residuals are not autocorrelated (Durbin-Watson Test, p-value: $0.65$). As already seen in the previous section using the correlation coefficients, the categories \cat{Layman} and \cat{Dependence on Time} positively correlate with \cat{Sociality} -- however, only \cat{Layman} is statistically significant ($p=0.00$). \cat{Operator} has a negative impact on \cat{Sociality} (but the result is statistically not significant with $p=0.12$). 

\begin{table}
{\footnotesize
 	 	\begin{tabularx}{\columnwidth}{Zrrrr} 
 	 	\toprule
        \tableheadline{Variable} & 
        \tableheadline{Coefficient} & 
        	\tableheadline{Std. Error} &
        	\tableheadline{t value} &
        	\tableheadline{$P(>|T|)$} \\
        \midrule
  (Intercept) & 0.1654 & 0.0742 & 2.23 & 0.0305 \\ 
  (Dependence on Time) & 0.1651 & 0.1185 & 1.39 & 0.1699 \\ 
  Layman & 0.2531 & 0.0778 & 3.25 & 0.0021 \\ 
  Operator & -0.0945 & 0.0605 & -1.56 & 0.1248 \\  
                \bottomrule
                \end{tabularx}
      }            
      \caption[Linear regression model to explain degree of sociality]{\label{tab:exp3_soc_model}Linear regression model to explain degree of sociality, Residual standard error: $0.0733$ on $48$ degrees of freedom, F-statistic: $5.675$ on $3$ and $48$ DF,  p-value: $0.00$, Adjusted R-squared:  $0.2157$}
\end{table}

When fitting and optimizing a linear regression model for the content categories (cf. \autoref{tab:exp3_soc_model2}), \cat{Finance \& Insurance} and \cat{Health \& Lifestyle} have a negative impact on \cat{Sociality} due to negative factors in the linear model ($-0.1331$ and $-0.1622$). Both values are statistically significant ($p=0.03$, $p=0.01$). \cat{Factual Knowledge \& News} has a positive impact (coefficient: $0.2727$) on a statistically significant level ($p=0.01$). While a negative correlation with \cat{Finance \& Insurance} can intuitively be explained, given the maturity, seriousness, and high personal impact of the domain, \cat{Health \& Lifestyle} and \cat{Finance \& News} are unexpected. \cat{Health \& Lifestyle} could be explained by the fact that people would like to consume passive information and have only a limited disposition to discuss individual problems with other users. The residuals of the model are normally distributed (Shapiro-Wilk normality test: W = $0.97$, p-value = $0.17$), studentized Breusch-Pagan test: p-value = $0.74$, Goldfeld-Quandt test: p-value = $0.11$) and no autocorrelation can be shown (Durbin-Watson Test, p-value = $0.20$).
  
\begin{table}
{\footnotesize
 	 	\begin{tabularx}{\columnwidth}{Zrrrr} 
 	 	\toprule
        \tableheadline{Variable} & 
        \tableheadline{Coefficient} & 
        	\tableheadline{Std. Error} &
        	\tableheadline{t value} &
        	\tableheadline{$P(>|t|)$} \\
        \midrule
(Intercept) & 0.2925 & 0.0125 & 23.40 & 0.0000 \\ 
  Entertainment & -0.0713 & 0.0434 & -1.64 & 0.1076 \\ 
  Finance \& Insurance & -0.1331 & 0.0580 & -2.30 & 0.0264 \\ 
  Health \& Lifestyles & -0.1622 & 0.0604 & -2.69 & 0.0101 \\ 
  Factual Knowledge \& News & 0.2727 & 0.1044 & 2.61 & 0.0122 \\ 
  Games & 0.0999 & 0.0605 & 1.65 & 0.1059 \\ 
  Ethics \& Phiosophy & -0.1110 & 0.0653 & -1.70 & 0.0961 \\                 \bottomrule
                \end{tabularx}
      }            
      \caption[Linear regression model to explain degree of sociality using content categories]{\label{tab:exp3_soc_model2}Linear regression model to explain degree of sociality using content categories, Residual standard error: $0.0711$ on $45$ degrees of freedom, F-statistic: $4.035$ on $6$ and $45$ DF,  p-value: $0.00$, Adjusted R-squared:  $0.2631$}
\end{table}


\begin{table}
{\footnotesize
 	 	\begin{tabularx}{\columnwidth}{Zrrrr} 
 	 	\toprule
        \tableheadline{Variable} & 
        \tableheadline{Coefficient} & 
        	\tableheadline{Std. Error} &
        	\tableheadline{t value} &
        	\tableheadline{$P(>|T|)$} \\
        \midrule
(Intercept) & 0.0970 & 0.0466 & 2.08 & 0.0427 \\ 
  Layman & 0.2413 & 0.0984 & 2.45 & 0.0178 \\ 
  Operator & 0.1139 & 0.0806 & 1.41 & 0.1637 \\ \bottomrule
                \end{tabularx}
      }            
      \caption[Linear regression model to explain degree of mobility]{\label{tab:exp3_mob_model}Linear regression model to explain degree of mobility, Residual standard error: $0.0978$ on $49$ degrees of freedom, F-statistic: $3.324$ on $2$ and $49$ DF,  p-value: $0.04$, Adjusted R-squared:  $0.0835$}
\end{table}

When analyzing the content categories, \cat{Ethics \& Philosophy} appears to impact mobility positively (but is not statistically significant, i.e. $p=0.06$), while \cat{Recreation} is suggested to be considered as a negative factor (\autoref{tab:exp3_mob_cont_model}). The model fulfills formal statistical requirements: the studentized Breusch-Pagan test and the Goldfeld-Quandt test did not reject the homoscedasticity hypothesis (p=$0.52$ and p=$0.69$) and the Durbin-Watson Test does not give evidence for autocorrelation in the residuals (p=$0.29$).

\begin{table}
{\footnotesize
 	 	\begin{tabularx}{\columnwidth}{Zrrrr} 
 	 	\toprule
        \tableheadline{Variable} & 
        \tableheadline{Coefficient} & 
        	\tableheadline{Std. Error} &
        	\tableheadline{t value} &
        	\tableheadline{$P(>|T|)$} \\
        \midrule
(Intercept) & 0.2113 & 0.0147 & 14.34 & 0.0000 \\ 
  Ethics \& Phiosophy & 0.1664 & 0.0875 & 1.90 & 0.0631 \\ 
  Recreation & -0.1643 & 0.0765 & -2.15 & 0.0368 
  \\ \bottomrule
                \end{tabularx}
      }            
      \caption[Linear regression model to explain degree of mobility using content categories]{\label{tab:exp3_mob_cont_model}Linear regression model to explain degree of mobility using content categories, Residual standard error: $0.0959$ on $49$ degrees of freedom, F-statistic: $4.468$ on $2$ and $49$ DF,  p-value: $0.02$ }
      
\end{table}

\section{Limitations}
\label{exp3:limitations}
The findings of the conducted experiment need to be interpreted carefully: the experiment covers only a limited sample of websites and it is not possible to guarantee that the randomly chosen URLs reflect the assigned content categories completely. In addition, the axes which were chosen to classify information are based on initial assumptions, but can not be considered exhaustive. The existence of other suitable axes is quite likely (e.g., degree of emotionality or degree of assurance).

\section{Conclusion}
The findings suggest differences in the degree of sociality and mobility for various content areas and other attribute types. The most obvious finding is that laymen positively correlate with fields that can be characterized by a large degree of sociality. \cat{Factual Information \& News} (or opinions on these topics), \cat{Games}, and \cat{Business Profession} show the highest correlation with \cat{Sociality}, while \cat{Health \& Lifestyle} is negatively correlated.

\bibliographystyle{IEEEtran}
\bibliography{literature}

\appendix
\onecolumn
\begin{table}[ht]
	\centering
    \includegraphics[origin=c,angle=0,width=0.84\textwidth]{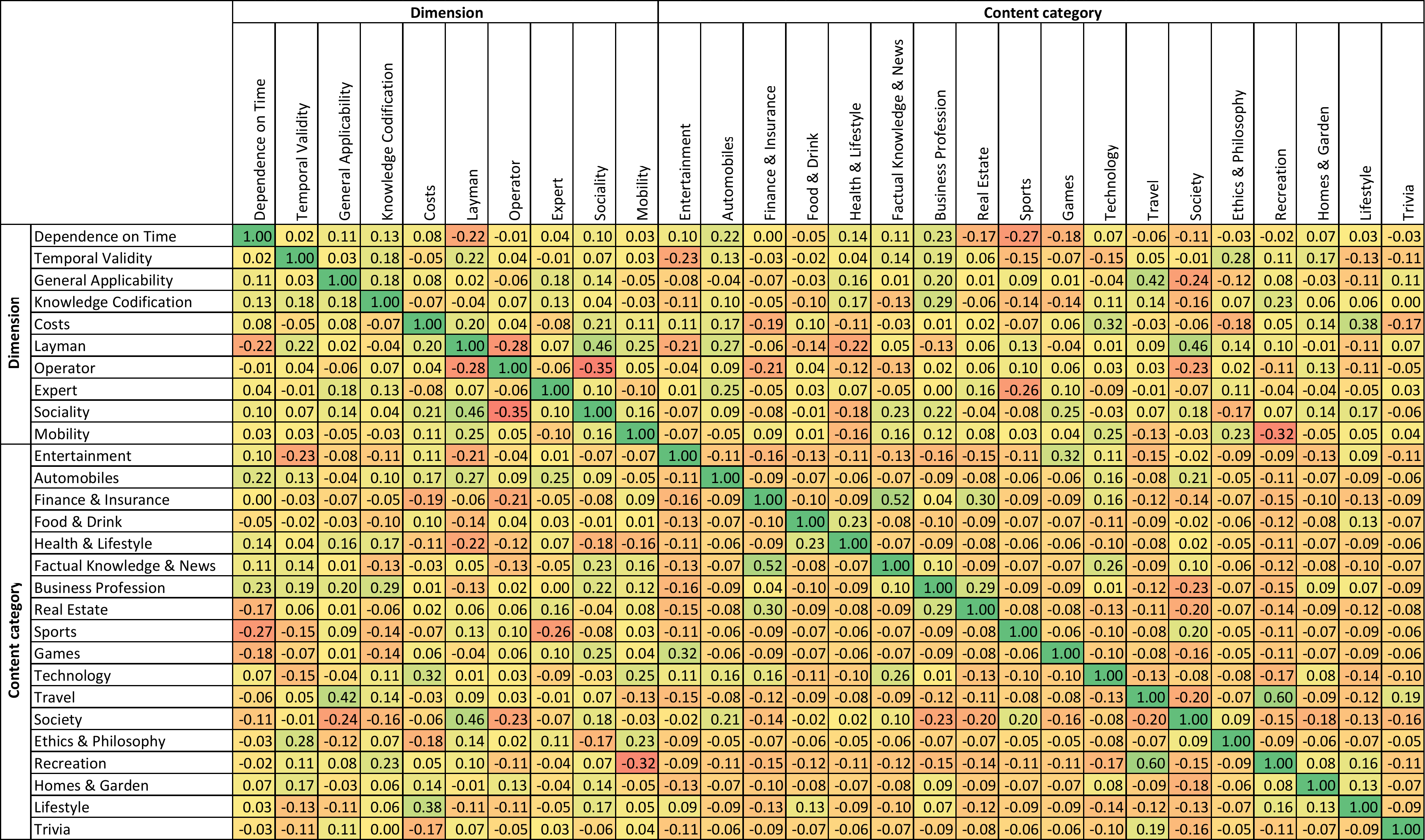}  
    \caption{\label{tab:ex3_spearman}Correlation between dimensions and content categories of information needs (Spearman's rho)}
\end{table}

\begin{table}[ht]
	\centering
    \includegraphics[origin=c,angle=0,width=0.84\textwidth]{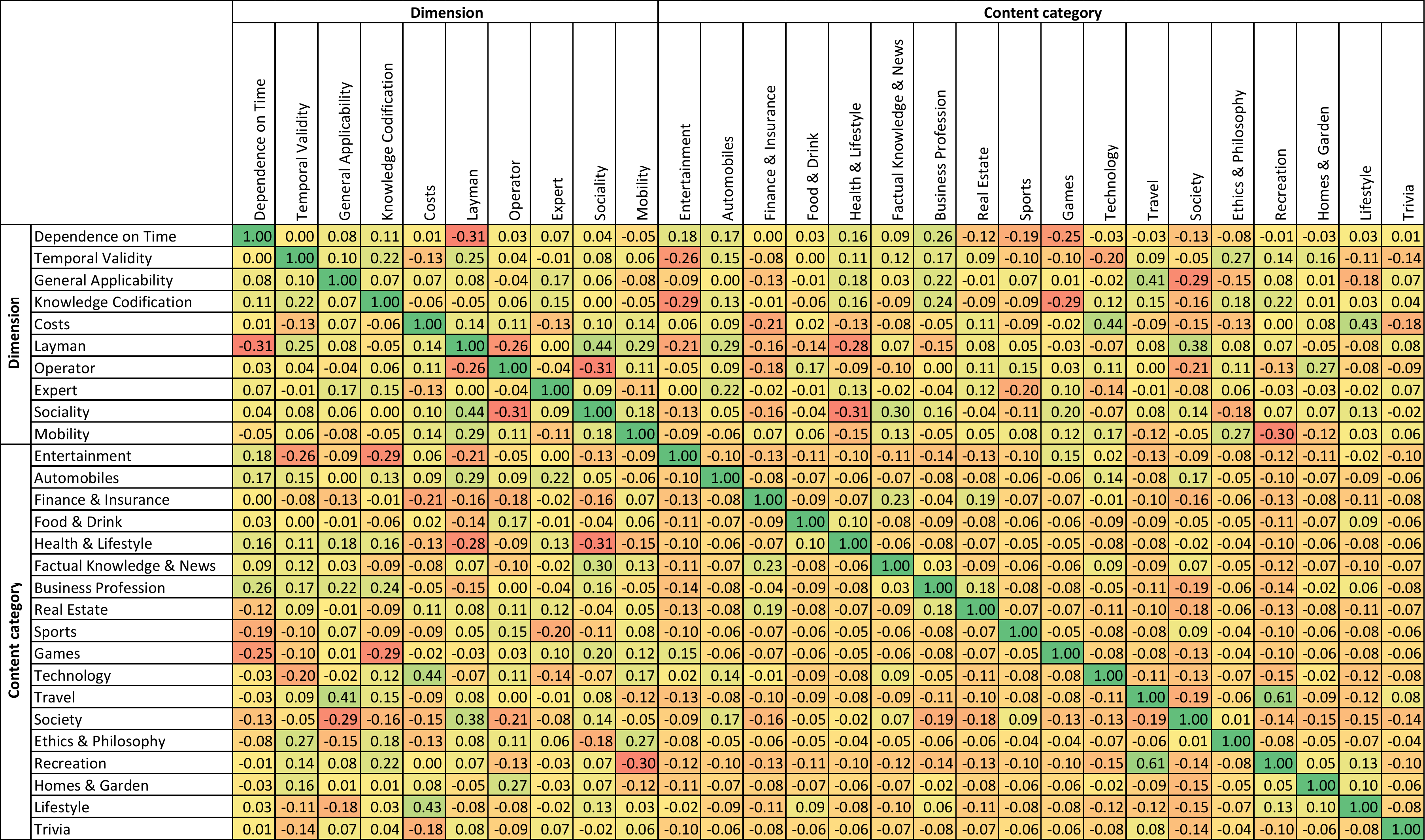}  
    \caption{\label{tab:ex3_pearson}Correlation between dimensions and content categories of information needs (Pearson's r)}
\end{table}



\end{document}